\documentclass[conference]{IEEEtran}
\IEEEoverridecommandlockouts
\usepackage[numbers,square,comma,compress,sort]{natbib}
\usepackage{booktabs} 

\usepackage{graphicx, dblfloatfix}
\usepackage{xcolor}
\usepackage{amsmath}
\usepackage{amsfonts}
\usepackage[ruled, linesnumbered]{algorithm2e}
\usepackage{enumitem}
 \usepackage{multirow}

\newcommand{\camready}[1]{\textcolor{black}{#1}}

\begin{document}
\title{Theoretical Analysis and Evaluation of NoCs \\ with Weighted Round-Robin Arbitration}

\author{Sumit K. Mandal$^1$, Jie Tong$^1$, Raid Ayoub$^2$, Michael Kishinevsky$^2$, Ahmed Abousamra$^2$, Umit Y. Ogras$^1$ \\
$^1$Dept. of ECE, University of Wisconsin-Madison; 
$^2$Intel Corporation, Hillsboro, OR
\vspace{-3mm}
\thanks{This work was supported by Strategic CAD Labs, Intel Corporation, USA.}
}

\maketitle



\begin{abstract}
\label{sec:abstract}
Fast and accurate performance analysis techniques are essential in early design space exploration and pre-silicon evaluations, including software eco-system development. 
In particular, on-chip communication continues to play an increasingly important role as the many-core processors scale up.
This paper presents the first performance analysis technique that targets networks-on-chip (NoCs) that employ weighted round-robin (WRR) arbitration.
Besides fairness, WRR arbitration provides flexibility in allocating bandwidth proportionally to the importance of the traffic classes, unlike basic round-robin and priority-based arbitration.
The proposed approach first estimates the \textit{effective service} time of the packets in the queue due to WRR arbitration.
Then, it uses the effective service time to compute the average waiting time of the packets.
Next, we incorporate a decomposition technique to extend the analytical model to handle NoC of any size.
\camready{The proposed approach achieves less than 5\% error while executing real applications and 10\% error under challenging synthetic traffic with different burstiness levels.}

\end{abstract}


\section{Introduction} 
\label{sec:introduction}

Networks-on-chip continue playing a central role as many-core processors start dominating the server~\cite{arafa2019cascade, doweck2017inside}
\camready{and deep learning~\cite{sun2021reliability, mandal2020latency, nabavinejad2020overview}}
market. As the commercial solutions scale up, the latency, area, and power consumption overheads of NoCs become increasingly crucial. 
Designers need analytical power-performance models to guide complex design decisions during the architecture development and implementation phases. After that, the same models are required by virtual platforms, commonly used to develop and evaluate the software ecosystem and applications~\cite{chiang2011qemu}. Hence, there is a strong demand for high-fidelity analytical techniques that accurately model fundamental aspects of industrial designs across all segments ranging from systems-on-chip to client and server systems.

NoCs can be broadly classified in terms of buffer usage as buffered and bufferless architectures~\cite{moscibroda2009case, daya2016quest,marculescu2008outstanding,dally2004principles}. Most early solutions adapted buffered techniques, such as wormhole and virtual-channel switching, where the packets (or their flits) are stored in intermediate routers. Area, latency, and energy consumption of buffers have later led to bufferless architectures, where the intermediate routers forward the incoming flits if they can and deflect otherwise. 
Bufferless NoCs save significant buffer area and enable ultra-fast, as low as single-cycle routing decisions~\cite{moscibroda2009case, daya2016quest}. Therefore, many industrial NoCs used in server and client architectures employ bufferless solutions to minimize the communication latency between the cores, last-level caches (LLC), and main memory~\cite{sodani2016knights, doweck2017inside}. These solutions give priority to the packets already in the network to enable predictable and fast communication while stalling the newly generated packets from the processing and storage nodes. However, buffer area savings and low communication latency come at the cost of the early onset of congestion. 
Indeed, the packets wait longer at the end nodes, and the throughput saturates faster when the NoC load increases. 
Moreover, all routers in the NoCs remain powered on, increasing the NoC power consumption.
Therefore, there is a strong need to address these shortcomings.

Buffered NoCs with virtual channel routers have been used more commonly in academic work and most recent industry standards~\cite{ogras2013modeling,pellegrini2020arm}. Shared buffering resources, such as input and output channels, require arbitrating among different requesters. For example, suppose that packets in different input channels request the same output channel. An arbiter 
needs to resolve the conflicts and grant access to \camready{one of the requesters to meet performance target}. The architectures proposed to date predominantly employ basic round-robin (RR) arbiter to provide fairness to all requesters~\cite{shin2002round, lee2009high, xiaopeng2006round}. Although the decisions are locally fair, the number of arbitrations a packet goes through grows with its path length. Hence, RR arbitration is globally \camready{unfair}.
More importantly, basic RR cannot provide preference to a particular input, which is typically desired since not all requests are equal. For example, data and acknowledgment packets can have higher priority than new requests to complete outstanding transactions, especially when the network is congested. 


WRR arbitration provides flexibility in allocating bandwidth proportionally to the importance of the traffic classes, unlike basic round-robin and priority-based arbitration. 
Each requester has an assigned weight, which is a measure of its importance.
A larger weight indicates that the requester is given more preference in arbitration.
Due to its generality, WRR arbitration has been employed in several NoC proposals in the literature~\cite{qian2010qos, heisswolf2012scalable, zeferino2003socin}.
Indeed, WRR arbitration enables higher throughput than RR arbitration~\cite{heisswolf2012scalable}. 
\textit{Despite its potential, WRR arbitration has not been analyzed theoretically, especially for large-scale NoCs.}
A large body of literature has proposed performance analysis techniques for buffered and bufferless NoCs since analytical models play a crucial role in fast design space exploration and pre-silicon evaluation~\cite{wang2010analytical, qian2015support, kiasari2012analytical, mandal2020analytical, mandal2020performance}.
In contrast, no analytical modeling technique has been proposed to date for NoCs with WRR arbitration. 
A formal analysis is required to understand the behavior of NoCs with WRR arbitration. At the same time, executable performance models are needed to guide many-core processor design and enable virtual platforms for pre-silicon evaluation.


This paper presents a fast, accurate, and scalable performance analysis technique for NoCs with WRR arbitration. 
To the best of our knowledge, it is the first performance analysis technique for NoCs with weighted round-robin arbitration.
Furthermore, the proposed technique supports bursty core traffic observed in real applications, which is typically ignored due to its complexity.
It first estimates the effective service time of the packets in the queue due to WRR arbitration.
Then, it utilizes the effective service time to obtain the average waiting time of the packets. 
We also propose a decomposition technique to extend the analytical model for any size of NoC.
Extensive experimental evaluations show that the proposed analytical model has less than 
\camready{5\% with real applications and 10\% error with synthetic traffic having different burstiness levels congesting the NoC.}

The major contributions of the work are listed below:
\begin{itemize}
    \item A novel performance analysis technique for NoCs that employ WRR arbitration,
    \item A decomposition technique to obtain a scalable analytical model for NoC of any size.
    \item Experimental evaluations with multiple NoC configurations with different traffic scenarios showing less than 5\% error \camready{for real applications}.
\end{itemize}

\section{Related Work and Novel Contributions} 
\label{sec:related_work}


Analytical models are required to estimate the NoC performance for fast design space exploration and pre-silicon evaluation. 
Multiple prior studies have proposed NoC performance analysis techniques with basic round-robin arbitration.~\cite{ogras2010analytical, fischer2013accurate, qian2015support}.
Authors in~\cite{ogras2010analytical} first construct a contention matrix between multiple flows in the NoC.
Then, the average waiting time of the packets corresponding to each flow is computed.
Support vector regression-based analytical model for NoCs is proposed in~\cite{qian2015support}. 
The analytical model proposed in~\cite{fischer2013accurate} estimates the mean service time of the flows with RR arbitration.
The estimated mean service time is used to find the average waiting time of the flows.
However, none of these techniques are applicable in the presence of both bursty traffic and WRR arbitration.

Analytical modeling of round-robin arbitration has also been studied outside NoC domain~\cite{boxma1987waiting, groenendijk1992performance, wang2010analytical}.
The techniques presented in~\cite{boxma1987waiting, groenendijk1992performance} incorporate a polling model to approximate the effective service time of a queue in the presence of RR arbitration.
However, none of these approaches are applicable when the input distribution to the queue is not geometric.
A Markov chain-based analytical model is proposed in~\cite{wang2010analytical} to account for bursty input traffic.
However, the technique is not scalable for a network of queues.
Moreover, none of these techniques are applicable for discrete-time queuing systems.
Since each transaction in NoC happens at discrete clock cycles, the analytical models need to incorporate discrete-time \camready{queuing systems}.
The major drawbacks of the prior approaches are summarized in Table~\ref{tab:prior_work}.

The basic round-robin arbitration cannot provide fairness when requesters have widely varying data rate requirements and priorities.
Therefore, weighted round-robin arbitration, i.e., WRR, has been used in on-chip communication architectures~\cite{qian2010qos, heisswolf2012scalable}.
Qian et al. compute delay bounds for different channels with different weights to assign appropriate weight to each input channel of the NoC~\cite{qian2010qos}.
They show that WRR delivers better quality of service than NoCs with strict priority-based arbitration.
Authors in~\cite{heisswolf2012scalable} propose a WRR-based scheduling policy.
The proposed technique assigns larger bandwidth to input channels with higher weights.
It achieves higher throughput compared to round-robin arbitration.
\textit{Although WRR has shown promise, no analytical modeling approach exists for NoCs with WRR to date.}

\begin{table}[t]
\centering
\vspace{-2mm}
\caption{\camready{Comparison} of prior research and our novel contribution.}
\vspace{-2mm}
\setlength\tabcolsep{2pt}
\begin{tabular}{|l|l|l|l|l|l|}
\hline
\textbf{Research}                                             & \textbf{Approach}                                                      & \textbf{WRR} & \textbf{\begin{tabular}[c]{@{}l@{}}Bursty\\ Traffic\end{tabular}} & \textbf{Scalable} & \textbf{\begin{tabular}[c]{@{}l@{}}Discrete\\ Time\end{tabular}} \\ \hline
Boxma et al.~\cite{boxma1987waiting}                                                  & Polling model                                                          & \textbf{No}            & No                                                                 & Yes                 & No                                                                \\ \hline
Wim et al.~\cite{groenendijk1992performance}                                                    & \begin{tabular}[c]{@{}l@{}}Extended\\ polling model\end{tabular}       & \textbf{No}            & No                                                                 & Yes                 & No                                                                \\ \hline
Wang et al.~\cite{wang2010analytical}                                                   & Markov chain                                                           & \textbf{No }           & Yes                                                                 & No                 & No                                                                \\ \hline
Fischer et al.~\cite{fischer2013accurate}                                                & Heuristic                                                              & \textbf{No}           & No                                                                 & Yes                 & No                                                                \\ \hline
\begin{tabular}[c]{@{}l@{}}Vanlerbergee\\ et al.~\cite{vanlerberghe2018analysis}\end{tabular} & \begin{tabular}[c]{@{}l@{}}Moment\\ generating function\end{tabular}   & \textbf{No}            & No                                                                 & No                 & Yes                                                                \\ \hline
\textbf{This work}                                            & \textbf{\begin{tabular}[c]{@{}l@{}}Queue\\ decomposition\end{tabular}} & \textbf{Yes}   & \textbf{Yes}                                                        & \textbf{Yes}        & \textbf{Yes}                                                       \\ \hline
\end{tabular}
\vspace{-4mm}
\label{tab:prior_work}
\end{table}

This paper presents the first performance analysis technique for NoCs with WRR arbitration. 
It fills an essential gap since WRR can address the shortcomings of priority-based bufferless \camready{NoC} architectures and the basic round-robin arbitration.
Furthermore, the proposed technique supports bursty traffic observed in real applications, which is typically ignored due to its complexity. 
Hence, it is a vital step towards comprehending the theoretical underpinnings of NoCs with WRR arbitration and enabling their deployment in industrial designs.


\section{Background and Overview}
\label{sec:background}


\begin{figure}[b]
	\centering
	\vspace{-6mm}
	\includegraphics[width=0.78\columnwidth]{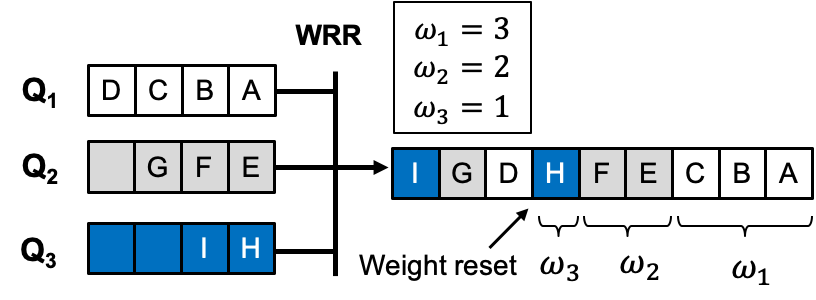}
	\vspace{-3mm}
	\caption{Illustration of weighted round-robin arbitration. \camready{`A' is served first, `I' last. In this example it is assumed that no new packets arrived until all prior packets (A-I) have been served.}}
	\vspace{-4mm}
	\label{fig:wrr_ex}
\end{figure}

\subsection{Weighted Round-Robin Arbitration}

This work uses weighted round-robin arbitration in the NoC routers. 
The basic
\camready{operating} principle of WRR arbitration is illustrated in 
Figure~\ref{fig:wrr_ex} for three traffic classes.
Packets from each class are first written to a dedicated input queue (also known as a channel).
Suppose there are $N$ input queues $Q_1, Q_2, \ldots, Q_N$.
WRR technique assigns $Q_i$ a positive integer weight denoted as $\omega_i \in \mathbb{Z}^+$ for $1 \leq i \leq N$. 
The WRR arbiter serves up to $\omega_i$ consecutive packets from $Q_i$ before moving to the
\camready{next} queue. 
If $Q_i$ has less than $\omega_i$ packets, then WRR serves $Q_i$ until it becomes empty. 
Then, the WRR arbiter serves the subsequent queues following the same principle. 
After a cycle is completed, the weights are reset to their initial values, as illustrated in Figure~\ref{fig:wrr_ex}, and the same arbitration cycle is repeated.
In NoCs with WRR arbitration, whenever two or more requesters compete for the same resource, they are arbitrated following WRR.
WRR arbitration can be used both for arbitrating different virtual channels and different ports in the network.


\subsection{Usage of the Proposed Performance Analysis Technique}

WRR arbitration is promising for NoCs since it can tailor the communication bandwidth to different traffic classes. Furthermore, it 
\camready{provides} end-to-end latency-fairness to different source-destination pairs, unlike basic round-robin and priority arbitration techniques. 
However, these capabilities come at the expense of a vast design parameter space. 
An $n\times m$ mesh with $P$-port routers has $n \times m \times P$ tunable weights, e.g., an 8$\times$8 2D mesh with 5-port routers would have 320 WRR weights. 
Due to this ample design space, the current practice is limited to assigning two weights to each router (e.g., one weight to local ports and another weight to packets already in the NoC).

The benefits of the proposed theoretical analysis are two-fold.
\textit{First}, it can enable accurate pre-silicon evaluations and design space exploration without time-consuming cycle-accurate simulations.
\textit{Second}, it can be used to find the combination of weights that optimizes the performance, i.e., to solve the optimization problem 
\camready{in the vast design space}
described in the previous paragraph. 
\textit{This paper focuses on constructing the proposed analysis technique and its evaluation against cycle-accurate simulation for two reasons.} 
First, the theoretical analysis is complex, and its evaluation deserves a dedicated treatment on its own.
Moreover, its application for solving optimization problems requires 
demonstration of its fidelity first.
Multi-objective optimization of the WRR weights using the proposed model is
one of our future research directions.

\begin{figure}[b]
	\centering
    \vspace{-5mm}
	\includegraphics[width=0.9\columnwidth]{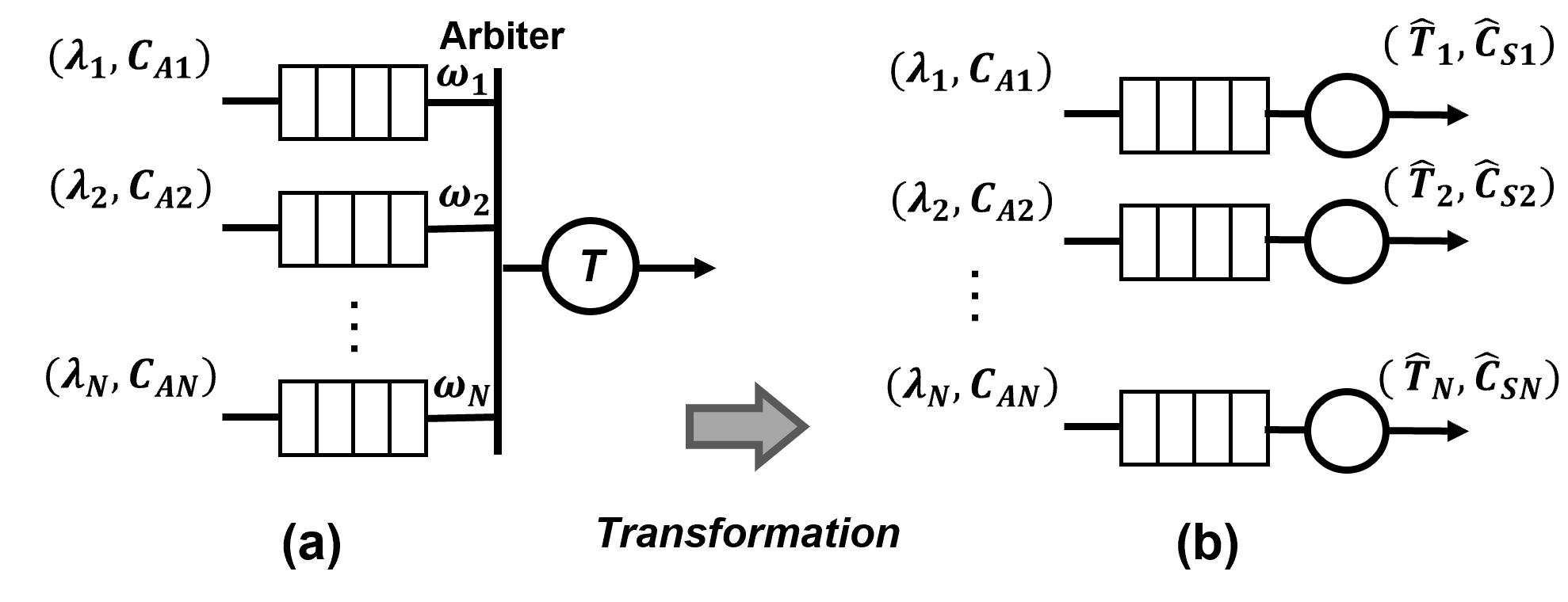}
	\vspace{-5mm}
	\caption{The original WRR arbiter with its traffic parameters (on the left) and a transformed WRR (on the right) that comprises fully decomposed queue nodes using our effective service time transformations.}
	\label{fig:canon}
	\vspace{-4mm}
\end{figure}

\section{Proposed Methodology and Approach} 
\label{sec:methodology}

Figure~\ref{fig:canon} shows a
weighted round-robin arbiter with $N$ queues.
The packets that belong to traffic class-$i$ are stored in queue $Q_i$. 
The corresponding average arrival rate of class-$i$ packets is 
\camready{denoted} by $\lambda_i$, as summarized in Table~\ref{tab:notation}. 
The burstiness of traffic class-$i$ can be captured by their squared
coefficient of variation \camready{of inter-arrival time}, denoted by $C_{Ai}$~\cite{kouvatsos1994entropy, qian2010qos}.
Finally, the weight assigned to class-$i$ is denoted as  $\omega_i \in \mathbb{Z}^+$ for $1 \leq i \leq N$, as illustrated in Figure~\ref{fig:canon}.

To provide a step-by-step derivation, Section~\ref{sec:rr} starts with a particular case of the proposed technique tailored to the basic round-robin arbitration, i.e., all weights are set to one. 
Then, Section~\ref{sec:wrr} extends the formulation to weighted round-robin arbitration.

\begin{table}[t]
\vspace{-3mm}
\setlength\tabcolsep{0.75pt}
\caption{List of the important parameters used in this work.}
\vspace{-1mm}
\label{tab:notation}
\centering
	\vspace{-1mm}
	\begin{tabular}{@{}ll@{}}
\hline \\[-1em]
$N$     &  Number of traffic classes to an arbiter  \\  \hline \\[-1em]
$\lambda_i$    & Injection rate of class-$i$   \\ \hline \\[-1em]
$\omega_i$     & Weight assigned to class-$i$   \\ \hline \\[-1em]
$T$  & Original mean service time of all traffic classes                                                                                                    \\ \hline \\[-1em]
$\widehat{T}_i$  & Mean value of effective service time of class-$i$                                                                                                    \\ \hline \\[-1em]
$\rho_i$, $\widehat{\rho}_i$        & \begin{tabular}[c]{@{}l@{}}  ($\rho_i =  \lambda_i T_i$) and ($\widehat{\rho}_i = \lambda_i \widehat{T}_i$) server utilizations of class-$i$                                            \end{tabular}                      \\ \hline \\[-1em]
$C_{Ai}$  & \begin{tabular}[c]{@{}l@{}} Squared coeff. of variation of inter-arrival time of class-$i$  \end{tabular} \\ \hline \\[-1em]
$C_{S}$    & \begin{tabular}[c]{@{}l@{}} Squared coeff. of variation of orig. service time (all classes)\end{tabular} \\ \hline \\[-1em]
$\widehat{C}_{Si}$    & \begin{tabular}[c]{@{}l@{}} Squared coeff. of variation of eff. service time of class-$i$\end{tabular} \\ \hline \\[-1em]
$C_A$     & Sq. coeff. of variation of inter-arrival time (merged traffic)   \\ \hline \\[-1em]
$C_D$     & Squared coeff. of variation of inter-departure time  \\ \hline \\[-1em]

$n_i$       & Mean queue occupancy of class-$i$                                                                                                      \\ \hline \\[-1em]
\begin{tabular}[c]{@{}l@{}}$p(n_i>$ $k|\widehat{T}_j)$ \end{tabular} &  \begin{tabular}[c]{@{}l@{}} Probability that queue occupancy of class-$i$ $>$ $k$ \\ given a mean effective service time of class-$j$ = $\widehat{T}_j$ \end{tabular} \\ \hline \\[-1em]
$W_i$       & Average waiting time of class-$i$                                                                                                      \\ \hline \\[-1em]
$\widehat{R}_i$     &   Mean effective residual service time of class-$i$ \\ \hline
\end{tabular} 
\vspace{-4mm}
\end{table}

\subsection{Analytical Model for \textbf{Basic} Round-Robin Arbitration}
\label{sec:rr}

In the basic round-robin arbitration, all weights are equal to one, i.e., $\omega_i = 1,~1 \leq i \leq N$. 
Each traffic class experiences an additional delay until the arbiter serves the \camready{head} packets in the other queues. 
The proposed technique 
has two steps:
\begin{enumerate} [leftmargin=*]
    \item The first step is to compute the first two moments of the \textit{effective service time} for each class-$i$ ($\hat T_i$, $\hat C_{Si}$) of the fully decomposed queue nodes illustrated in Figure~\ref{fig:canon}. 

    \item In the second step, we use the transformed effective service times ($\hat T_i$, $\hat C_{Si}$, $1 \leq i \leq N$) to find the total waiting time (including the queuing delay) of each traffic class.
\end{enumerate}
For the clarity of notation and illustration, the derivation below assumes the original \camready{two moments of service time $(T, C_{S})$ are the same across all traffic classes.} 


\subsubsection{\textbf{Mean effective service time of RR ($\widehat{T}_i$)}}
The effective service time accounts for the delay experienced by the packets at the head of each queue. This delay includes its own service time and the service time of the packet at the head of other queues since the round-robin arbiter serves them one by one. 
When \camready{all queues have packets,}
class-$i$ packets will be served every $N \times T$ cycles, 
i.e., they will wait for $(N-1) \times T$ cycles after being served before winning the arbitration again. 
In general, a traffic class will only contribute to the extra service time if it has a packet waiting for service.
Thus, \camready{assuming packet arrival of different classes are independent,} we can express the mean effective service time as:
\vspace{-2mm}
\begin{align} \label{eq:t_1_rr}  \nonumber
    \widehat{T}_i &= T + \sum_{\substack{j=1, j \neq i}}^{N}p(n_i > 0|\widehat{T}_i)p(n_j > 0|\widehat{T}_i)T \\
    &= T + Tp(n_i > 0|\widehat{T}_i) \hspace{-3mm} \sum_{\substack{j=1, j \neq i}}^{N}p(n_j > 0|\widehat{T}_i),1 \leq i \leq N
\end{align}
%
\camready{where $p(n_j > 0|\widehat{T}_i)$ denotes the probability that the occupancy of class-$j$ is greater than zero given a mean effective service period of class-$i$ is $\widehat{T}_i$, for $1 \leq i,j \leq N$.
We approximate $p(n_j > 0|\widehat{T}_i)$ using Little's law as $\lambda_j \widehat{T}_i$ following the busy-cycle approach~\cite{boxma1987waiting}.
Little's law gives the expected number of class-$j$ packets that arrive during the period of $\widehat{T}_i$, 
which approximates the probability that the occupancy of class-$j$ packets is greater than zero during this period.}
\camready{To ensure that the probability term does not exceed one, we approximate $p(n_j > 0|\widehat{T}_i)$ as $min(1, \lambda_j \widehat{T}_i)$.}
Therefore, Equation~\ref{eq:t_1_rr} can be rewritten as:
\vspace{-2mm}
\begin{align} \label{eq:t_1_rr_2}
    \widehat{T}_i &= T + T min(1, \lambda_i \widehat{T}_i) \sum_{\substack{j=1 \\ j \neq i}}^{N} min(1, \lambda_j  \widehat{T}_i),~1 \leq i \leq N 
\end{align}
%
We solve Equation~\ref{eq:t_1_rr_2} through a simple iterative approach described in Algorithm~\ref{algo:t_hat_RR}.
First, Equation~\ref{eq:t_1_rr_2} is solved by removing the nonlinearity introduced by the $min$ operation (lines 4 and 5).
We consider the smaller solution out of the two solutions, \camready{since we observed through experiments that the smaller solution is more accurate}.
The solution is used as the initial estimate of the iterative approach.
Then, this estimate is plugged into Equation~\ref{eq:t_1_rr_2} within an iteration loop to obtain a better estimate (line 7).
The iterations continue until the change in the effective service time is within a user-provided $Tolerance$ value, set to $0.01$ in this work.
Algorithm~\ref{algo:t_hat_RR} typically takes less than ten iterations due to the quadratic nature of Equation~\ref{eq:t_1_rr_2}, which takes only a few microseconds to compute.
After the iterations are completed, Algorithm~\ref{algo:t_hat_RR} returns the effective service time $\widehat{T}_i$.

\begin{algorithm}[t]
\small
\caption{Basic round-robin arbitration: Effective service time ($\widehat{T}_i$) computation.} \label{algo:t_hat_RR}
\SetAlgoLined
\SetNoFillComment
\textbf{Input:} Injection rate of each class ($\lambda$), service time ($T$), number of classes ($N$), Convergence $Tolerance$ \\
\textbf{Output:} Effective service time of each class ($\widehat{T}$) \\

\For {i = 1:N} {
%
Find smaller root of the quadratic \camready{equation derived from} Equation~\ref{eq:t_1_rr_2}: \\
$ T+ T \lambda_i (\widehat{T}_i)^2 \sum_{\substack{j=1, j \neq i}}^{N} \lambda_j -  \widehat{T}_i = 0$ \\
$\delta = Tolerance, k=1$ \\
\While {$\delta \geq Tolerance$} {
\hspace{-2mm} $\widehat{T}_i^{k+1} \gets T + Tmin(1,\lambda_i \widehat{T}_i^k) \sum_{\substack{j=1 \\ j \neq i}}^{N} min(1, \lambda_j  \widehat{T}_i^k)$ \\
$\delta = \widehat{T}_i^{k+1} - \widehat{T}_i^k$ \\
$k \gets k+1$\\
}
}
\vspace{-1mm}
\end{algorithm}


\subsubsection{\textbf{Coefficient of variation of effective service time for RR ($\widehat{C}_{Si}$)}}

This section derives $\widehat{C}_{Si}$ by leveraging the conservation of work principle and property of equal residual service time of individual classes \cite{boxma1987waiting}. 
We leverage these properties to compute the \camready{mean effective residual service time of class-$i$ ($\widehat{R}_i$)}.
Then, $\widehat{R}_i$
is used to calculate $\widehat{C}_{Si}$ using the relation found in~\cite{kouvatsos1994entropy}:
%
\vspace{-4mm}
\begin{equation} \label{eq:cs_i_rr}
    \widehat{C}_{Si} = \frac{1}{\widehat{\rho}_i} \bigg( \frac{2\widehat{R}_i}{\widehat{T}_i} + 1 - C_{Ai} - \widehat{\rho}_i \bigg)
\end{equation}
where $\widehat{\rho}_i = \lambda_i \widehat{T}_i$ is the effective server utilization. In addition, we show that leveraging the conservation of work principle gives our model the capability of handling bursty traffic. 

We exploit the \textit{conservation of work principle}, 
\camready{stating that} the same amount of packets are served regardless of how the bandwidth is divided between the traffic flows \camready{assuming all packets have same moments of service time}. 
Due to this fundamental principle, the total queue occupancy $n_{sum}$ is invariant under the arbitration policy. Furthermore, this sum can be computed with near-perfect accuracy using 
\camready{maximum entropy-based} analytical queuing model of multi-class given in~\cite{kouvatsos1994entropy}.
One strong aspect of this model is that it can take bursty traffic that conforms to General Geometric distribution (GGeo).
By leveraging this model, we can embed the capability of handling bursty traffic into our analytical model. 
The total queue occupancy ($n_{sum}$) can be calculated as:
\vspace{-3mm}
\begin{align} \label{eq:n_sum_me}
    n_{sum} = \frac{1}{2}\sum_{i=1}^{N} \Bigg(\rho_i (C_{Ai} - 1) + \frac{\displaystyle\sum_{k=1}^{N} \frac{\lambda_i}{ \lambda_k} \rho_k^2 (C_{Ak} + C_{S} )}{1 - \sum_{k=1}^{N} \rho_k} \Bigg)
\end{align}
using the parameters summarized in Table~\ref{tab:notation}. We do not delve into the derivation of this equation for clarity since it is not our major focus. The detailed derivation can be found in the seminal work by Kouvatsos et al.~\cite{kouvatsos1994entropy}.

Next, we derive $n_{sum}$ using the first two moments of transformed effective service time.
The occupancy of each queue $Q_i$ can be expressed using Little's law as: $n_i = \lambda_i W_i$, where $W_i$ is the average waiting time of class-$i$. Therefore, we obtain:
\vspace{-3mm}
\begin{equation} \label{eq:n_sum}
    n_{sum} = \sum_{i=1}^{N}  \lambda_i W_i
\end{equation}
The average waiting time $W_i$ consists of two components.
The first one is the waiting time in the queue till the packet reaches the head of the queue ($\widehat{W}_i$), i.e. until all other class-$i$ packets leave the server completely.
The second component accounts for the waiting time at the head of the queue when the server is busy serving other classes (class-$j$, $j \neq i$). 
This component is
the additional time captured by the effective service time $\hat T_i$, i.e., the difference between the effective and original service times: $\Delta T_i = \widehat{T}_i - T$.
%
Furthermore, the waiting $\widehat{W}_i$ is expressed as a function of the residual time~\cite{bertsekas1992data}. Hence,
\vspace{-2mm}
\begin{equation} \label{eq:w_i}
    W_i = \frac{\widehat{R}_i}{1-\lambda_i \widehat{T}_i} + \Delta T_i,~for~1 \leq i \leq N
\end{equation}
We leverage the property that all individual \camready{mean effective residual service time of classes} are equal \cite{boxma1987waiting}; hence we write ($R = \widehat{R}_i, 1 \leq i \leq N$). 
Thus, we can rewrite the total occupancy by plugging 
Equation~\ref{eq:w_i} into Equation~\ref{eq:n_sum} and \camready{then} solve for the residual time $R$:
\vspace{-2mm}
\begin{align} \nonumber
    n_{sum} &=  \sum_{i=1}^{N} \lambda_i \Big( \frac{R}{1 - \lambda_i \widehat{T}_i} + \Delta T_i \Big) \\ 
    \implies R &= \Big( n_{sum} - \sum_{i=1}^{N} \lambda_i \Delta T_i \Big) \Big( \sum_{i=1}^{N}  \frac{\lambda_i}{1 - \lambda_i \widehat{T}_i} \Big) ^ {-1} \label{eq:R_rr}
\end{align}
Then, we can compute the coefficient of variation $\widehat{C}_{Si}$ by substituting the residual time $R$ from Equation~\ref{eq:R_rr} in Equation~\ref{eq:cs_i_rr}.


\subsubsection{\textbf{Average waiting time of RR ($W_i$)}}

Finally, we compute the mean waiting time of individual classes 
by plugging $\widehat{T}_i$ and $\widehat{C}_{Si}$ found for RR arbitration into \camready{Equation~\ref{eq:cs_i_rr} and} Equation~\ref{eq:w_i} as:
\vspace{-2mm}
\begin{equation} \label{eq:w_i_rr_final}
    W_i = \frac{0.5\widehat{T}_i (\widehat{\rho}_i - 1 + C_{Ai} + \widehat{\rho}_i \widehat{C}_{Si})}{1-\lambda_i \widehat{T}_i} + \Delta T_i, ~1 \leq i \leq N
\end{equation}

\subsection{Analytical Model for \textbf{Weighted} Round-Robin Arbitration}
\label{sec:wrr}

This section extends the analytical model constructed in Section~\ref{sec:rr} to WRR using the same steps. 
It first describes the derivation of the first two moments of effective service time.
Then, it will use these moments to find the average waiting time.



\subsubsection{\textbf{Mean effective service time of WRR ($\widehat{T}_i$)}}

In WRR, the weight \camready{of each class} can be larger than or equal to one, i.e., $\omega_i \geq 1,~1 \leq i \leq N$.
To illustrate the generalization,
we start with a system of two queues, where $\omega_1 > 1$ and $\omega_2~=~1$. Then, we generalize it to $N$ queues with arbitrary weights. 

\textit{\camready{Mean effective service time of \textbf{class-$1$}} with $\omega_1 > 1$ and $\omega_2 = 1$}:
Since $\omega_1 > 1$, a batch of up to $\omega_1$ number of class-$1$ packets are served without interruption.
Let $\widehat{T}_i^b$ be the total service time of $\omega_1$ class-$1$ packets.
When there is a class-$2$ packet in $Q_2$, the next batch of $\omega_1$ class-$1$ packet will wait.
Therefore, we can express $\widehat{T}_1^b$ as:
\vspace{-2mm}
\begin{align} \label{eq:tc1}
    \widehat{T}_1^b = \omega_1 T + p(n_1 > 0|\widehat{T}_1^b) p(n_2 > 0|\widehat{T}_1^b)T
\end{align}
Note that this equation generalizes Equation~\ref{eq:t_1_rr} to arbitrary $\omega_1$ for two queues.
Using Little's law and the methodology described in Section~\ref{sec:rr}, we approximate $\widehat{T}_1^b$ as:
\vspace{-2mm}
\begin{equation} \label{eq:t_1_b}
    \widehat{T}_1^b = \omega_1 T + \frac{T}{\omega_1} min(1, \lambda_1 \widehat{T}_1^b) min(1, \lambda_2 \widehat{T}_1^b) 
\end{equation}
%

\textit{\camready{Mean effective service time of \textbf{class-$2$}} with $\omega_1 > 1$ and $\omega_2 = 1$}:
The batch size of class-$2$ packet is one in this case since $\omega_2 = 1$.
They need to wait in the queue for a maximum of $\omega_1$ class-$1$ packets. 
Therefore, $\widehat{T}_2$ is expressed as:
\vspace{-2mm}
\begin{equation} \label{eq:t_2_hat_wrr}
    \widehat{T}_2 = T + \sum_{i=0}^{\omega_1-1} p(n_1 > i|\widehat{T}_2) p(n_2 > 0|\widehat{T}_2) T
\end{equation}
We can approximate $p(n_1 > i|\widehat{T}_2)$ as $\frac{1}{i+1} p(n_1 > 0|\widehat{T}_2)$ to obtain:
\vspace{-4mm}
\begin{align} \nonumber
    & \widehat{T}_2 = T + \sum_{i=0}^{\omega_1-1} \frac{1}{i+1} p(n_1 > 0|\widehat{T}_2) p(n_2 > 0|\widehat{T}_2) T \\ \label{eq:t_final_wrr}
                  &= T + T min \Big(1, \sum_{i=0}^{\omega_1-1} \frac{1}{i+1} \lambda_1 \widehat{T}_2 \Big) min(1, \lambda_2 \widehat{T}_2)    
\end{align}
To get to the final form of Equation~\ref{eq:t_final_wrr} above, we approximate the conditional probabilities based on~\cite{boxma1987waiting}.

\noindent\textit{Generalization to $N$ queues and arbitrary weights:}
In general, there are $N$ classes with weights $\omega_i, 1 \leq i \leq N$. 
Consider the class-$i$ packets with 
the total service time $\widehat{T}_i^b$ for
\camready{a batch of} $\omega_i$ packets. 
The first part of the effective service time will be due to the \textit{packets \camready{of} same class}. This effect is captured in our two-queue illustration for class-$1$ packets in Equation~\ref{eq:tc1}. 
The second part of the effective service time will be due to the \textit{packets \camready{of} other classes}. This effect is captured in our two-queue illustration for class-$2$ packets in Equation~\ref{eq:t_2_hat_wrr}. 
By combining them, we can express the generalized effective service time as:
\vspace{-5mm}
\begin{equation} \label{eq:t_i_hat_genl}
    \widehat{T}_i^b = \omega_i T + p(n_i > 0|\widehat{T}_i^b) \hspace{-2mm} \sum_{\substack{j=1, j \neq i}}^{N} \hspace{-1mm} \Big( \sum_{k=0}^{\omega_j-1}  p(n_j > k|\widehat{T}_i^b) \Big)T 
\end{equation}
Note that this equation reduces to  Equation~\ref{eq:tc1} for $i=1, N=2, \omega_2=1$, 
and to Equation~\ref{eq:t_2_hat_wrr} for $i=2, N=2, \omega_2=1$. 

Finally, we can obtain the generalized effective service times by replacing $p(n_i > 0)$ with $min(1, \lambda_i \widehat{T}_i^b)$ as follows:
\vspace{-1mm}
\begin{align} \label{eq:t_hat_genl}
    \widehat{T}_i^b = \omega_i T + \frac{T}{\omega_i} min(1, \lambda_i \widehat{T}_i^b) & \sum_{\substack{j=1 \\ j \neq i}}^{N} min \Big(1, \sum_{k=0}^{\omega_j-1} \frac{1}{k+1} \lambda_j \widehat{T}_i^b \Big) \nonumber \\
    \widehat{T}_i &= \frac{\widehat{T}_i^b}{\omega_i} 
\end{align}
Due to the non-linearity introduced by the $min$ operation, 
we compute $\widehat{T}_i$ using the iterative Algorithm~\ref{algo:t_hat} just like the basic round-robin case. 
The quadratic nature of Equation~\ref{eq:t_hat_genl} enables a fast convergence within $10$ iterations under $5\mu$s.

\begin{algorithm}[t]
\small
\caption{Obtaining effective service time ($\widehat{T}_i$) of weighted round-robin} \label{algo:t_hat}
\SetAlgoLined
\SetNoFillComment
\textbf{Input:} Injection rate of each class ($\lambda$), WRR weights ($\omega_i$), service time ($T$), number of classes ($N$) \\
\textbf{Output:} Effective service time of each class ($\widehat{T}$) \\

\For {i = 1:N} {
%
\hspace{-2mm} Find smaller root of the quadratic \camready{equation derived from} Equation~\ref{eq:t_hat_genl} for $\widehat{T}_i^b$:\\
$ (\widehat{T}_i^b)^2 \frac{T}{\omega_i} \lambda_i \sum_{\substack{j=1 \\ j \neq i}}^{N} \sum_{k=0}^{\omega_j-1} \frac{1}{k+1} \lambda_j -  \widehat{T}_i^b  + \omega_i T = 0$ \\
$\delta = Tolerance, k=1$ \\
\While {$\delta \geq Tolerance$} {
$Z = \sum_{\substack{j=1 \\ j \neq i}}^{N} min \Big(1, \sum_{k=0}^{\omega_j-1} \frac{1}{k+1} \lambda_j \widehat{T}_i^{b, k} \Big)$ \\
$\widehat{T}_i^{b, k + 1} \gets \omega_i T + \frac{T}{\omega_i} min(1, \lambda_i \widehat{T}_i^{b, k}) Z$ \\ 
$\delta = \widehat{T}_i^{b, k+1} - \widehat{T}_i^{b,k}$ \\
$k \gets k+1$\\
}
$\widehat{T}_i = \frac{\widehat{T}_i^b}{\omega_i}$
}
\end{algorithm}

\begin{figure*}[b!]
	\centering
	\vspace{-3mm}
	\includegraphics[width=2\columnwidth]{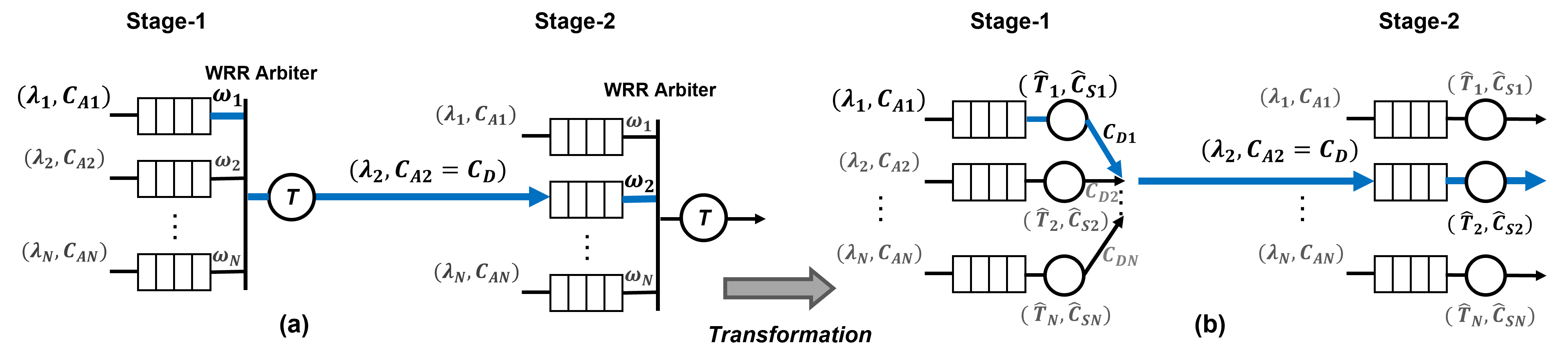}
	\vspace{-4mm}
	\caption{Illustration of extending the proposed analysis to multiple stages. Only two consecutive stages are shown for clarity. The departure statistics at a given stage become the arrival statistics of the subsequent stage. {The blue line denotes that the class which wins the arbitration goes to the next stage.}}
	\vspace{-16mm}
	\label{fig:decomp}
\end{figure*}


\subsubsection{\textbf{Coefficient of variation of effective service time of WRR
($\widehat{C}_{Si}$)
}\protect\footnote{We use the same notation for RR and WRR arbitration since WRR results are generalizations of basic RR expressions.}}
In the basic round-robin case, the residual service time $R_i$ accounts for the second-order moment of the service time, i.e., $C_{Si}$.
However, \camready{effective mean residual service time} of different classes are not necessarily equal to each other for arbitrary weights, which is the case in WRR arbitration.  
%
The \camready{mean effective residual service time} of class-$i$ packets is expressed as:
%
\vspace{-3mm}
\begin{equation}
 \camready{\widehat{R}_i = 0.5\widehat{T}_i (\widehat{\rho}_i - 1 + C_{Ai} + \widehat{\rho}_i \widehat{C}_{Si})}
\vspace{-1mm}
\end{equation}
where $\widehat{T}_i$ is found \camready{from Algorithm~\ref{algo:t_hat} based on}
Equation~\ref{eq:t_hat_genl}. 
Note that the residual service time depends on the first two moments of service time,
\camready{which are the average service time $\widehat{T}_i$ and the coefficient of variation $\widehat{C}_{Si}$, respectively.  
To approximate $\widehat{C}_{Si}$ for WRR,}
we leverage its monotonically decreasing behavior as a function of the corresponding weight. This behavior is expected since a larger weight corresponds to fewer arbitration stages, 
\camready{decreasing}
the effective service time variation.  
\camready{This observation indicates that the coefficient of variation for round-robin arbitration is an upper bound of coefficient of variation for weighted round-robin arbitration.}
This upper bound, $\widehat{C}_{Si}(RR)$ for class-$i$, is computed using Equation~\ref{eq:cs_i_rr}.
Given this, we first apply linear approximation
\camready{to calculate the $\widehat{C}_{Si}$ under WRR as
$\alpha \frac{\widehat{C}_{Si} \text{(RR)}}{\omega_i^2}$,}
note we use the squared value of $\omega_i$ since $\widehat{C}_{Si}$ is the square of coefficient of variation as defined in Table~\ref{tab:notation}.
We add the parameter $\alpha$ to tighten the approximation.
Next, we recall Equation~\ref{eq:n_sum_me} and replace $C_{Sk}$ by \camready{$\alpha \frac{\widehat{C}_{Sk}(RR)}{\omega_k^2}$} to obtain:
\vspace{-4mm}
\begin{align} \label{eq:n_sum_cs} 
    n_{sum} \hspace{-1mm} = \hspace{-1mm} \frac{1}{2} \hspace{-1mm} \sum_{i=1}^{N} \hspace{-1mm} \Bigg(\widehat{\rho}_i (C_{Ai} - 1) +
    \frac{\displaystyle \hspace{-1mm} \sum_{k=1}^{N}  (\frac{\lambda_i}{\lambda_k} ) \widehat{\rho}_k^2 (C_{Ak} \hspace{-1mm} + \hspace{-1mm} \alpha \frac{\widehat{C}_{Sk}(RR)}{\omega_k^2})}{1 - \sum_{k=1}^{N} \widehat{\rho}_k}  \hspace{-1mm} \Bigg)
\end{align}
The left hand side of the Equation~\ref{eq:n_sum_cs} ($n_{sum}$) is obtained by applying Equation~\ref{eq:n_sum_me}.
Therefore, $\alpha$ is the only unknown in Equation~\ref{eq:n_sum_cs}. 
Once we obtain $\alpha$ \camready{(by solving Equation~\ref{eq:n_sum_cs})}, we compute
\camready{$\widehat{C}_{Si}$ under WRR as $\alpha \frac{\widehat{C}_{Si} \text{(RR)}}{\omega_i^2}$.}
%
This expression captures traffic burstiness addressing one of the significant drawbacks of the prior work.
%
%


\subsubsection{\textbf{Average waiting time of WRR ($W_i$)}}

So far, we obtained first two moments of effective service time, i.e., $\widehat{T}_i$ (Equation~\ref{eq:t_hat_genl}) and
\camready{$\widehat{C}_{Si}$.}
We obtain the average waiting time under WRR arbitration by plugging them to Equation~\ref{eq:w_i_wrr}:
%
\vspace{-1mm}
\begin{equation} \label{eq:w_i_wrr}
    W_i = \frac{0.5\widehat{T}_i (\widehat{\rho}_i - 1 + C_{Ai} + \widehat{\rho}_i \widehat{C}_{Si})}{1-\lambda_i \widehat{T}_i} + \Delta T_i 
\end{equation}

\subsection{Estimation of end-to-end latency}
\vspace{-2mm}

The previous section described the canonical model for WRR arbitration, where all traffic classes go through a single arbiter.
The packets in NoCs go through a sequence of WRR arbiters while traveling from their sources towards destinations.
For example, an 8$\times$8 2D mesh has 64 routers with at least one arbiter \camready{per output queue} in each of them.
Figure~\ref{fig:decomp}(a) illustrates an example with two stages, 
where the flow that wins the first stage is routed to the second one.
The major challenge in modeling multiple stages is obtaining the inter-departure distribution, which becomes the arrival flow at the subsequent stages. 
{For instance, the coefficient of variation of arrival flow 2 in the second stage is equal to the coefficient of variation of the departing flow from stage-1 ($C_{D}$) in Figure~\ref{fig:decomp}(a).}
{We first find the squared co-efficient of variation of inter-departure time for each traffic class ($C_{Di}$):
\vspace{-2mm}
\begin{equation} \label{eq:cd}
    C_{Di} \hspace{-1mm} = \hspace{-1mm} \rho_i^2 (\widehat{C}_{Si} + 1) \hspace{-0.5mm} + \hspace{-0.5mm} ( 1 - \rho_i ) C_{Ai} \hspace{-0.5mm} + \hspace{-0.5mm} \rho_i (1 - 2 \rho_i), \forall 1 \leq i \leq N
\vspace{-2mm}
\end{equation}
Next, we apply the decomposition technique~\cite{pujolle1986solution} to obtain the departure distribution at each stage ($C_D$).
\vspace{-2mm}
\begin{equation} \label{eq:merged_ca_rho}
    C_D = \frac{\sum_{i=1}^{N} \lambda_i C_{Di}}{\sum_{i=1}^{N} \lambda_i}
\end{equation}
The departure distribution ($C_D$) is the arrival distribution to the next stage ($C_{A2}$).
}
For a given source-destination pair, these calculations are performed at each arbitration stage on the path of the packets.

\begin{algorithm}[t]
\small
\caption{Estimation of end-to-end latency} \label{algo:end_to_end}
\SetAlgoLined
\SetNoFillComment
\textbf{Input:} NoC size ($n \times m$), injection rates and co-efficient of variation of inter-arrival times for all classes ($\lambda, C_A$),  Service time ($T$), WRR weights\\
\textbf{Output:} Average NoC \hspace{-1mm} ($\bar L$) and source--dest. latencies ($L_{sd}$)\\

Number of routers $N_p \gets n \times m$, 
$\lambda_{sum} \gets 0, L_{sum} \gets 0$ \\

\For {s = 1:$N_p$} {

\For {d = 1:$N_p$} {


$W_{sd} = compute\_waiting\_time (s, d)$  \\




$L_{sd} = W_{sd} + free\_packet\_delay (s, d)$ \\

$L_{sum} \gets L_{sum} + L_{sd}$, $\lambda_{sum} \gets \lambda_{sum} + \lambda_{sd}$

}
}
%
%
$\bar{L} = \frac{L_{sum}}{\lambda_{sum}}$
\SetKwProg{Fn}{function}{}{end}

\Fn{compute\_waiting\_time($s, d$) }{

$S \gets$ The number of stages for the flow with source $s$ and destination $d$, $W_{sd} \gets 0$ \\

\For {j = 1:S} {


   \hspace{-3mm} $W_j =$ Waiting time at stage $j$ using Equation~\ref{eq:w_i_wrr}\\

\hspace{-3mm} $W_{sd} \gets W_{sd} + W_j$

\hspace{-3mm} $C_D = $ Coefficient of variation using Equation~\ref{eq:merged_ca_rho} \\
\hspace{-3mm} Use $C_D$ as the $C_A$ in the next stage
}

return $W_{sd}$

}
\end{algorithm}

Algorithm~\ref{algo:end_to_end} presents a step-by-step procedure to obtain end-to-end latency of a given NoC with WRR arbitration and deterministic routing.
The input to the algorithm is the NoC size, injection rate, and coefficient of variation of inter-arrival time of all classes, service time of the queues, and weights assigned to each arbiter.
The output of the algorithm is the latency of each source-destination pair ($L_{sd}$) as well as the average latency $\bar L$.

Algorithm~\ref{algo:end_to_end} first computes the waiting time of each source-destination pair ($W_{sd}$).
The function $compute\_waiting\_time$ in lines 12--22 describes the procedure for this computation.
The procedure starts with finding the number of stages ($S$) for the flow from source $s$ to destination $d$.
At each stage, it computes the average waiting time using Equation~\ref{eq:w_i_wrr} (line 16).
Next, it adds the waiting time at the current state to the cumulative waiting time found so far (line 17).
Then, it computes the coefficient of variation of inter-departure time ($C_D$) using Equation~\ref{eq:cd} and $C_D$ is used as the coefficient of variation of inter-arrival time in the next stage (lines 18--19).
Finally, the procedure returns the total waiting time from source $s$ to destination $d$ ($W_{sd}$) (line 21). 
After retrieving the total waiting times, Algorithm~\ref{algo:end_to_end} adds the free packet delay ($t_{sd}$), i.e., the latency due to the links and router micro-architecture, which can be found using the number of hops and the router pipeline depth~\cite{ogras2006s,dally2004principles}.
Then, end-to-end latencies and flow rates are accumulated (line 8).
Finally, the average latency of the network is obtained by computing the weighted average of the latency of all source-destination pairs (line 11).
\camready{We note that different WRR arbiters in the network can be assigned different arbitration weights.}



\section{Experimental Results}
\label{sec:experimental_results}

\vspace{-1mm}
\subsection{Experimental Setup}
\vspace{-1mm}

This section presents a thorough evaluation of the proposed analytical modeling approach under various traffic scenarios and NoC configurations.
We employ geometric and bursty traffic distributions in addition to real application traces feeding 1$\times$8, 6$\times$6, and 8$\times$8 NoCs (note that the typical size of state-of-the-art industrial NoCs is 6$\times$6~\cite{sodani2016knights, doweck2017inside}).
The synthetic traffic simulations use 100\% last-level cache (LLC) hit and 100\% LLC miss scenarios, commonly used corner cases.
Furthermore, applications from SPEC CPU 2017~\cite{bucek2018spec} \camready{and PARSEC~\cite{bienia2008parsec}} benchmark suits demonstrate the effectiveness of the proposed technique in a broader range of scenarios.
The latency results of our analytical models are compared against an in-house cycle-accurate simulator, which is calibrated with an industrial simulator\camready{~\cite{ogras2012energy}}.
All simulations run for 200K cycles to reach a steady state and have 20K cycles \camready{of} warm-up.
\camready{The source code of the simulator and the analytical model are publicly released at~\cite{WRR_models}.}

\begin{figure}[t]
	\centering
	\vspace{-6mm}
	\includegraphics[width=0.8\columnwidth]{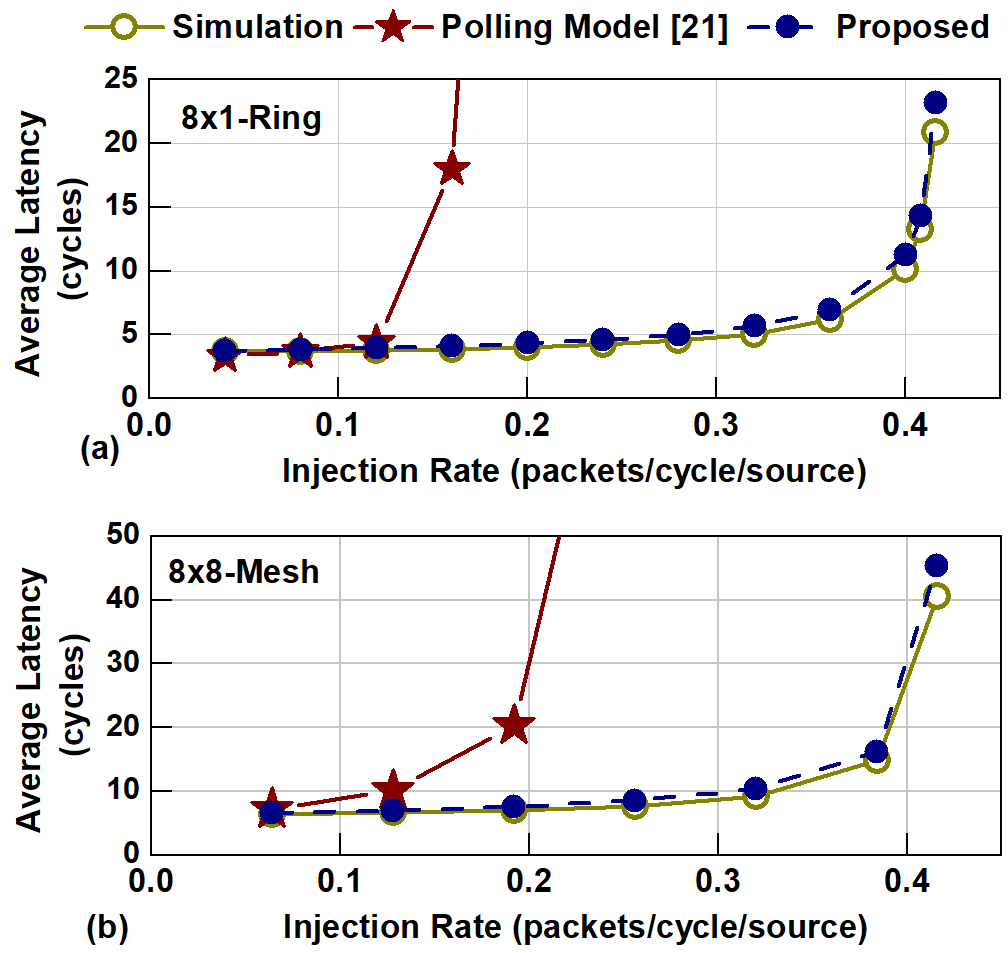}
	\vspace{-3mm}
	\caption{Verification of the analytical model for basic round-robin with (a) 8$\times$1 and (b) 8$\times$8 NoC.}
	\vspace{-6mm}
	\label{fig:basic_RR}
\end{figure}

\vspace{-1mm}
\subsection{Results with Basic Round-Robin Arbitration} \vspace{-1mm}
\label{sec:expt_basic_rr}

We first evaluate the proposed analytical model under basic round-robin arbitration.
Figure~\ref{fig:basic_RR} depicts representative average end-to-end latency comparisons between 
\camready{analytical model}
and the cycle-accurate simulations for 8$\times$1 ring and 8$\times$8 mesh NoC. The average injection rates for both NoC configurations are varied until the NoC becomes highly congested, following geometric distribution.
On average, the proposed technique incurs only 5\% error for 8$\times$1 ring and 7\% error for 8$\times$8 mesh. 
We also compare the proposed analytical model with a polling-based model for round-robin arbitration~\cite{groenendijk1992performance}.
The polling-based model grossly overestimates the latency both for 8$\times$1 and 8$\times$8 NoC.
A comprehensive summary for other NoC sizes and traffic patterns is presented in Section~\ref{sec:summary_expt}.
These results demonstrate that the proposed approach is accurate for difference NoC, traffic, and WRR parameters.

\vspace{-1mm}
\subsection{Results with Weighted Round-Robin Arbitration}\vspace{-1mm}
\label{sec:expt_wrr}

\begin{figure}[t]
	\centering
	\vspace{-6mm}
	\includegraphics[width=0.85\columnwidth]{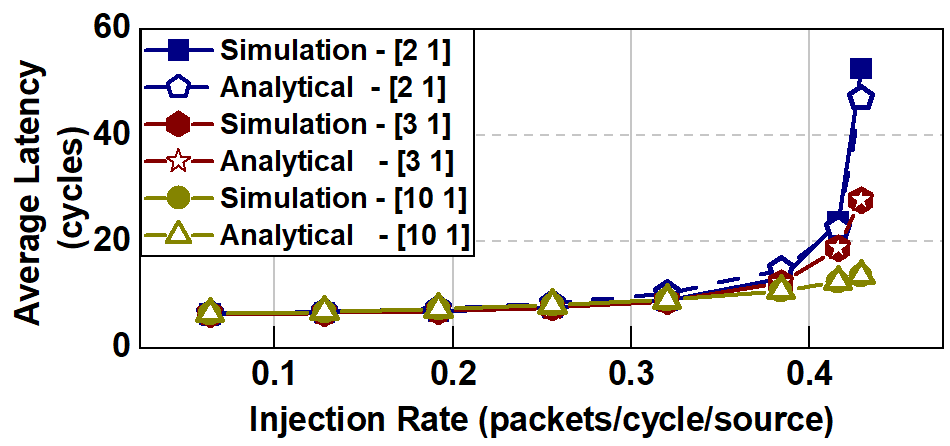}
	\vspace{-3mm}
	\caption{Verification of the analytical model for weighted round-robin with $8\times8$ NoC. [x y] denotes that the WRR weights associated with channels connected to the internal routers is x and traffic channels of external input to the NoC is y.}
	\vspace{-6mm}
	\label{fig:WRR}
\end{figure}

The most important aspect of the proposed technique is considering WRR arbitration.
Figure~\ref{fig:WRR} shows the average end-to-end latency comparison of our analytical model against cycle-accurate simulations \camready{with}
an 8$\times$8 mesh for three different weight configurations.
{The proposed analytical model incurs on average 8\% error when the WRR weights associated with the traffic channels of external input to the NoC and channels connected to internal routers are 1 and 2, respectively.
The average error slightly increases to 9\% when the arbitration weights of the packets in the NoC increase to 3.} This configuration decreases the network congestion by \camready{providing}
higher priority to the packets already in the NoC. Hence, the average waiting time decreases. 
Since there are no other analysis approaches for WRR, this section does not provide comparisons to state-of-the-art.

\begin{table*}[b]
\vspace{-5mm}
\begin{minipage}{.5\linewidth}
\setlength\tabcolsep{0.75pt}
\centering
\caption{Summary of results for synthetic applications with \textbf{100\% hit}. \label{tab:synth_hit}}
\vspace{-2mm}
\begin{tabular}{|c|c|c|c|c|c|c|c|c|}
\hline
\multirow{3}{*}{Topo.} & \multicolumn{4}{c|}{$\omega_{ring} =  1, \omega_{src} = 1$}                   & \multicolumn{4}{c|}{$\omega_{ring} = 3, \omega_{src} = 1$}                   \\ \cline{2-9} 
                          & \multicolumn{2}{c|}{$p_{burst} = 0$} & \multicolumn{2}{c|}{$p_{burst} = 0.3$} & \multicolumn{2}{c|}{$p_{burst} = 0$}  & \multicolumn{2}{c|}{$p_{burst} = 0.3$} \\ \cline{2-9} 
                          & $\lambda = 0.1$  & $\lambda = 0.3$ & $\lambda = 0.1$   & $\lambda = 0.3$  & $\lambda = 0.1$ & $\lambda = 0.3$ & $\lambda = 0.1$   & $\lambda = 0.3$  \\ \hline
8$\times$1              & 1.4\%             & 11\%           & 3.6\%                   &   7.8\%                & 1.5\%            & 13\%             & 9\%                & 11\%              \\ \hline
6$\times$6              & 5.5\%             & 7.2\%            & 7.4\%              & 11\%              & 5.2\%            & 11\%             & 5.9\%              & 12\%              \\ \hline
8$\times$8              & 2.6\%             & 7.8\%            & 5.2\%              & 10\%              & 3.5\%            & 11\%             & 4.8\%              & 7.2\%             \\ \hline
\end{tabular}
\end{minipage}%
    \begin{minipage}{.5\linewidth}
\setlength\tabcolsep{0.75pt}
\centering
\caption{Summary of results for synthetic applications with \textbf{100\% miss}. \label{tab:synth_miss}}
\vspace{-2mm}
\begin{tabular}{|c|c|c|c|c|c|c|c|c|}
\hline
\multirow{3}{*}{Topo.} & \multicolumn{4}{c|}{$\omega_{ring} =  1, \omega_{src} = 1$}                   & \multicolumn{4}{c|}{$\omega_{ring} = 3, \omega_{src} = 1$}                   \\ \cline{2-9} 
                          & \multicolumn{2}{c|}{$p_{burst} = 0$} & \multicolumn{2}{c|}{$p_{burst} = 0.3$} & \multicolumn{2}{c|}{$p_{burst} = 0$}  & \multicolumn{2}{c|}{$p_{burst} = 0.3$} \\ \cline{2-9} 
                          & $\lambda=0.1$  & $\lambda=0.2$ & $\lambda=0.1$   & $\lambda=0.2$  & $\lambda=0.1$ & $\lambda=0.2$ & $\lambda=0.1$   & $\lambda=0.2$  \\ \hline
8$\times$1              & 4.3\%             & 4.6\%           & 5.3\%                   & 5.7\%                  & 4.4\%            & 4.6\%             & 4.5\%                & 4.8\%              \\ \hline
6$\times$6              & 5.0\%             & 5.4\%            & 5.5\%              & 6.0\%              & 5.0\%            & 5.4\%             & 5.0\%              & 5.8\%              \\ \hline
8$\times$8              & 7.1\%             & 8.0\%            & 7.4\%              & 7.6\%              & 7.4\%            & 13\%             & 8.1\%              & 9.0\%             \\ \hline
\end{tabular}
\end{minipage} 
\vspace{-18mm}
	\end{table*}

\vspace{-2mm}
\subsection{Results with Bursty Traffic and Real Applications} \label{sec:expt_bursty}
\vspace{-1mm}

\begin{figure}[t]
	\centering
	\vspace{-5mm}
	\includegraphics[width=0.85\columnwidth]{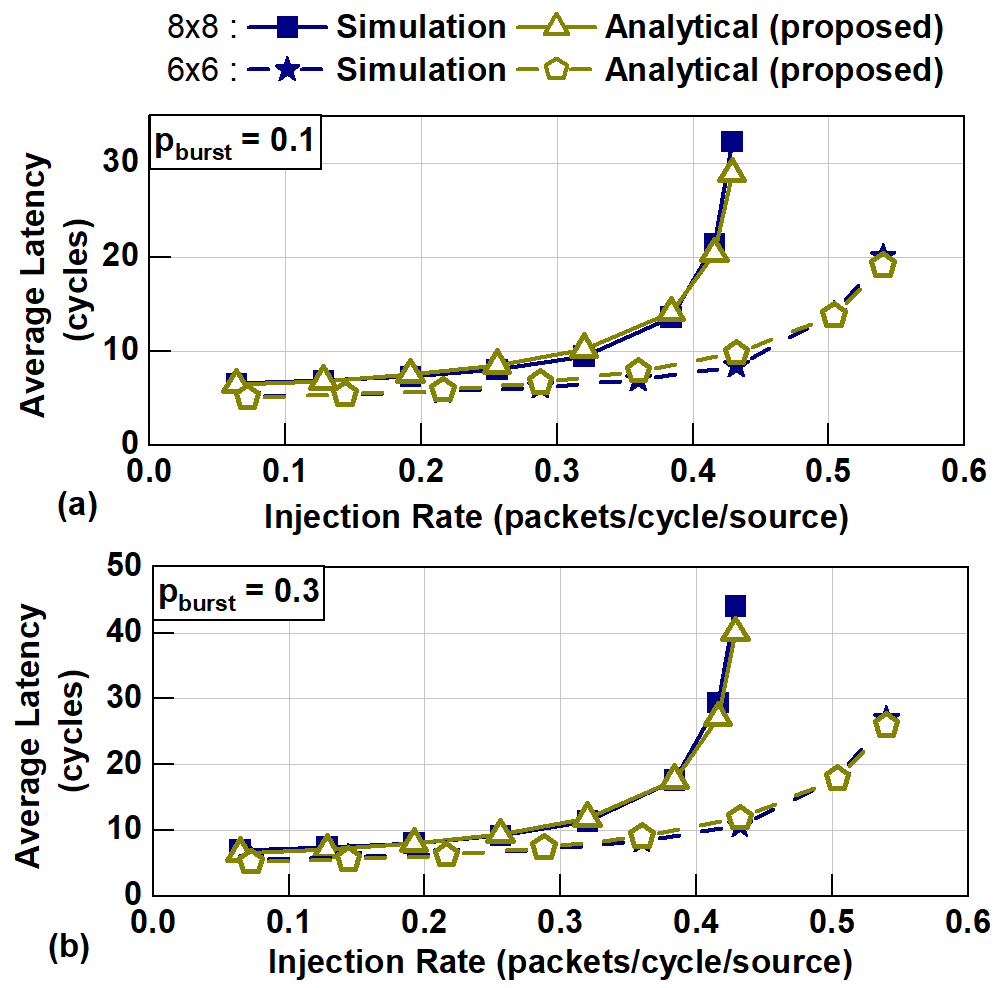}
	\vspace{-3mm}
	\caption{Verification of analytical model for bursty traffic with (a) $p_{burst}=0.1$ and $p_{burst}=0.3$.}
	\vspace{-5mm}
	\label{fig:burst}
\end{figure}

This section evaluates the proposed model in the presence of bursty traffic.
We define probability of burstiness as $p_{burst}$ of the general geometric distribution \cite{kouvatsos1994entropy}. Increasing $p_{burst}$ denotes increasing burstiness in the traffic.
We note that no other prior work considered weighted round-robin arbitration together with bursty traffic.
Figure~\ref{fig:burst} compares the end-to-end
\camready{average} latency results against cycle-accurate simulations for 6$\times$6 and 8$\times$8 mesh when the arbitration weights associated with the packets in the NoC and external packets arrivals are 3 and 1, respectively.
When the burst probability is $p_{burst}=0.1$ (Figure~\ref{fig:burst}(a)), 
the average modeling errors for 6$\times$6 and 8$\times$8 mesh are 7\% and 4\%, respectively.  
When the burst probability increases to $0.3$ (Figure~\ref{fig:burst}(b)), the 
average errors become 6\% and 5\%, respectively.   

\begin{figure}[t]
	\centering
	\vspace{-5mm}
	\includegraphics[width=0.85\columnwidth]{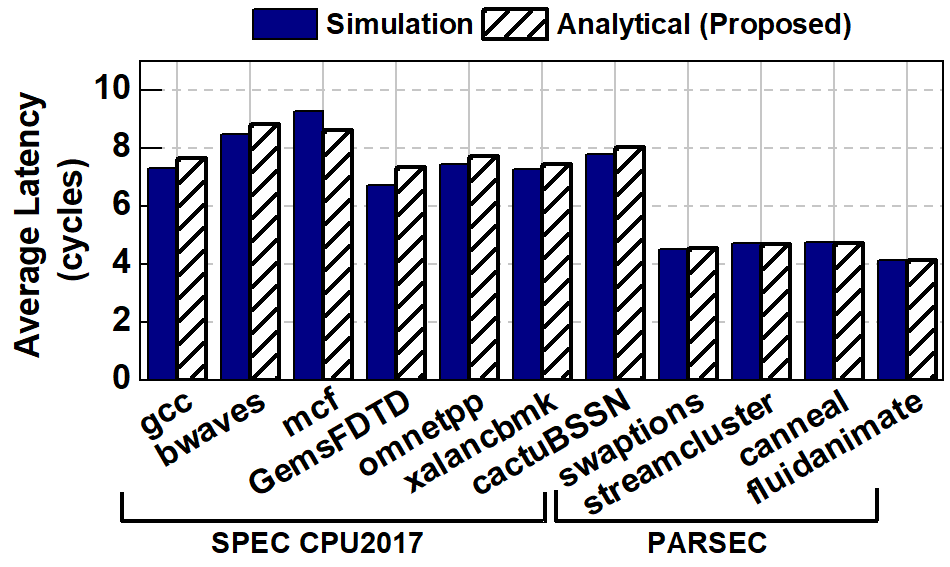}
	\vspace{-3mm}
	\caption{\camready{Verification against real applications.}}
	\vspace{-6mm}
	\label{fig:real_app}
\end{figure}

Finally, we evaluate the proposed analysis technique while running applications from the SPEC CPU2017~\cite{bucek2018spec}
\camready{and PARSEC
benchmark~\cite{bienia2008parsec}}.
\camready{The applications from SPEC CPU2017 benchmarks} show burstiness in the range of $p_{burst}=0.1 - 0.6$.
\camready{The applications from PARSEC benchmark are 16-threaded and do not show any burstiness.}
Figure~\ref{fig:real_app} compares the analysis results against cycle-accurate simulations on an 8$\times$8 NoC. 
The WRR weights of packets already in the NoC are set to 3, while new packets from cores have a lower priority set with a weight of 1. 
We observe that our proposed analysis technique consistently achieves a modeling error of less than 5\% for these applications.
These results demonstrate that the proposed analysis technique can accurately model the end-to-end NoC latency under WRR arbitration and bursty traffic.

\vspace{-1mm}
\subsection{Summary of the Evaluation Results} \label{sec:summary_expt}
\vspace{-1mm}

This section summarizes the accuracy of our performance analysis technique systematically for different NoC sizes, WRR weights, and traffic burstiness. 
To capture the most commonly observed corner cases, we provide results for
two extreme cases -- 100\% LLC hit and 100\% LLC miss traffic.
In both cases, cores send packets to each LLC at an equal rate.

Table~\ref{tab:synth_hit} summarizes the comparisons between analysis and simulations for 100\% LLC hit.
When $p_{burst} = 0$ and the traffic load is moderate ($\lambda=0.1$), the modeling error ranges from 1.4\% to 5.5\% considering both RR and WRR arbitration. 
Even when the traffic load increases congesting the NoC ($\lambda=0.3$), the modeling error remains below 11\% for all configurations. 
A higher level of burstiness to $p_{burst} = 0.3$ increases the NoC load. Consequently, the error range increases 3.6\%--9.0\% for the moderate traffic load ($\lambda=0.1$).  
\camready{Increasing the load further congests the NoC ($\lambda=0.3$) pushing the worst-case error only to}
13\%, which is acceptable since practical systems hardly operate this load due to congestion control mechanisms.


Finally, Table~\ref{tab:synth_miss} summarizes the modeling error of the proposed technique for 100\% LLC miss. 
In this case, the missed core requests are forwarded to memory controllers (one in 8$\times$1 NoC and two in 6$\times$6 and 8$\times$8 NoCs).
Hence, the memory controllers become hotspots.
Despite the high degree of skewness,  
the error for moderate NoC loads ranges from 4.3\% to 8.1\% for all burstiness levels, WRR weight configurations, and NoC sizes. 
Even when the NoC is pushed to congestion, the modeling error remains under 9.0\% for all configurations, except for $8 \times 8$ NoC with WRR weights 3 and 1. 
The error in this specific case is 13\%; the error for $8 \times 1$ and $6 \times 6$ NoCs are 4.6\% and 5.4\%, respectively.

\vspace{-1mm}
\subsection{Execution Time of the Proposed Analysis Technique}
\vspace{-1mm}
We implemented the proposed technique in C++ to obtain end-to-end latency.
The computational complexity of the proposed model is $O(n^2 \times m^2 \times p)$, for $n \times m$ NoC ($n \geq m$) and WRR arbiters with maximum $p$ ports. 
The execution time as a function of NoC sizes \camready{with 3 output ports per router} are summarized in Table~\ref{tab:exe_time}.
We observe that even for $32 \times 32$ NoC, the analytical model takes only 45.07 ms to execute, while the simulation time of the $8 \times 8$ NoC is in the order of minutes.
These results show that the proposed analysis technique is lightweight and provides four orders of magnitude speed-up compared to cycle-accurate simulations.

\begin{table}[h]
\centering
\vspace{-3mm}
\caption{Analysis on execution time of the proposed model.} \label{tab:exe_time}
\vspace{-2mm}
\begin{tabular}{|l|l|l|l|l|l|}
\hline
NoC size       & $4 \times 4$ & $6 \times 6$ & $8 \times 8$ & $16 \times 16$ & $32 \times 32$ \\ \hline
Exe. time (ms) & 0.56         & 1.41         & 2.25         & 13.52          & 45.07          \\ \hline
\end{tabular}
\vspace{-2mm}
\end{table}

\section{Conclusion and Future Work} \vspace{-1mm}
\label{sec:conclusion}

This paper presented a scalable performance analysis technique for NoCs with WRR arbitration. 
WRR arbitration is promising since it enables latency fairness and custom bandwidth allocation to meet the requirements of individual traffic flows, unlike basic round-robin and priority-based arbitrations.
The proposed technique handles bursty traffic and provides high accuracy for both ring topologies used in client systems and large 2D mesh topologies used in servers.
Extensive evaluations show the proposed approach achieves less than 
\camready{5\% error while executing real applications and 10\% error under challenging synthetic traffic with different burstiness levels.}
Hence, it can be used for fast and accurate design space exploration as well as pre-silicon evaluations.
One of our future directions is finding the WRR weights to meet individual traffic flows' bandwidth and latency requirements.



\bibliographystyle{unsrt}
{\small{
\bibliography{main}
}
}
\end{document}